# Viscoelastic behavior at high volume fraction of magnetorheological suspensions in the presence of discontinuous shear thickening

Volkova Olga
*UMR7010, Inphyni,*
*University Côte d'Azur, CNRS*
*17 rue Julien Lauprêtre,06200 Nice, France*
Olga.VOLKOVA@unice.fr

Grasselli Yan
*UMR7010, Inphyni,*
*University Côte d'Azur, SKEMA Business School*
*60 rue Dostoievski – CS30085*
*06902 Sophia Antipolis, France*
yan.grasselli@skema.edu

Bossis Georges
*UMR7010, Inphyni,*
*University Côte d'Azur,CNRS*
*17 rue Julien Lauprêtre,06200 Nice, France*
Georges.BOSSIS @unice.fr

*Abstract*— **This paper presents a study of the viscoelastic behavior of suspensions of carbonyl iron particles coated with a superplastifier molecule and dispersed in ethylene glycol. The experiments were done in plate-plate configuration with the rheometer MCR502 of Anton Paar. The discontinuous shear thickening (DST) transition manifests by a large jump of the storage and loss modulus. For low magnetic fields (H<50kA/m) the increase of moduli is at least of two orders of magnitude compared to the one induced by the magnetic field in the absence of DST. The behavior of the critical strain and the critical stress where the transition happens are presented for different frequencies and fields and a qualitative interpretation of the observed behavior is presented.**

***Keywords— Magnetorheological suspension, Viscoelasticity, Storage modulus, Discontinuous shear thickening***

## I. INTRODUCTION

The viscosity of a shear thickening fluid (STF) can increase by several orders of magnitude if an external force is applied on the liquid so it can be used in dampers to absorb shock (Zhang et al., 2008) and in bullet proof vest (Wei et al., 2025). On the other hand, magnetorheological fluids (MR) have the advantage of being active fluids whose viscosity can be controlled by a magnetic field with numerous applications in robotics, dampers, torque transmission etc. It is then appealing to use a shear thickening fluid as a carrier fluid in a dispersion of magnetic carbonyl iron particles to produce a MRSTF fluid. The rheology of each component of these MRSTF has been extensively studied and it is useful to recall the main findings corresponding to these two kinds of fluids. For STF fluids we shall distinguish shear thickening and discontinuous shear thickening (DST). Both behaviours are related to a change of the structure of the suspension under the action of external forces, in our case either hydrodynamic or magnetic, which in turn induces an increase of viscosity. In the case of suspensions of solid particles, they are often slightly aggregated at rest because they stay in a potential well, determined by the balance of attractive forces like Van der Waals or gravity forces and repulsive forces such as Debye-Huckel forces. This network of particles forms a gel which can be deformed elastically and then destroyed above a threshold stress- the yield stress, after the suspension begins to flow and the effective viscosity decreases along with the increase of the average distance between the particles. This is the liquid state where the viscous forces dominate the elastic ones. Still increasing the shear forces, the opposite phenomenon can appear when the Peclet number: $Pe=F_h/F_p$ is larger than unity. Here $F_h \propto \eta_s \dot{\gamma} a^2$ is the hydrodynamic force and $F_p$ the interparticle force deriving from a potential or the thermodynamic force ~kT/a; $\eta_s$ is the viscosity of the suspension at a given shear rate $\dot{\gamma}$ and a, the radius of the particles. In this situation, where the hydrodynamic forces dominate, the stationary sate is the one where all particles gather in aggregates where the lubrication forces dominate and stick the particles together. During this stage we can observe shear thickening due to the formation of transient aggregates with particles evolving mainly under the action of lubrication forces (Egres and Wagner, 2005). At still higher Peclet the distance between the particles decreases until their surfaces enter in frictional contact when the liquid layer between the particles has disappeared. This state can progress rapidly until a percolated network of frictional contacts occupy the whole space and link the walls of the cell, provoking a jump of viscosity (Barnes, 1989; Franks et al., 2000; Frith et al., 1996; Laun et al., 1991; Lootens et al., 2004). This transition is called discontinuous shear thickening and has been the subject of a growing interest from the last decade both theoretically (Wyart and Cates, 2014), experimentally (Comtet et al., 2017; Fall et al., 2015; Neuville et al., 2012) and with the help of numerical simulations (Mari et al., 2014; Seto et al., 2013). A more recent paper (Morris, 2025) emphasizes the link between dilatancy, particle pressure and DST. To summarize we can generally consider four steps in the response of a suspension of particles subjected to a shear flow. The first one is the gel state, which disappears above the yield stress. The second one is a shear thinning behavior when Pe<1 which is limited by a first critical stress $\sigma_{ST}$ where the viscosity begins to increase. The third step is the shear thickening state where the particles form aggregates whose particle surfaces are still separated by a lubrication layer until a critical stress $\sigma_{DST}$ is reached where the repulsive barrier becomes too weak to prevent a frictional contact between the particles. At this stage we have a percolation of the frictional contacts followed by a jump of viscosity. The fourth step corresponding to $\sigma > \sigma_{DST}$ often exhibits a new shear thinning domain, but experiments are questionable because it depends on the experimental conditions used for the measurement of the viscosity and on the wall slip. Actually, either the experiment is done under controlled stress i.e. the torque applied by a motor is raised continuously, and, in this case, the shear rate decreases during the jump and begins to oscillate around a constant value (Bossis et al., 2022; Frith et al., 1996; Laun, 1994) so the viscosity is still rising proportionally to the applied stress or the experiment is done under controlled shear rate and the jump of viscosity is associated with a jump of stress followed by a constant stress at higher shear rates corresponding to a decreasing viscosity, but this behavior could be due to a wall slip (Larsen et al., 2014). Along with the jump of viscosity due to the DST transition several authors have shown that the shear modulus or the complex viscosity behaves in the same way in oscillatory regime with the equivalence between the shear rate in steady sate and the product $\gamma.\omega$ of the strain amplitude times the angular frequency (Prabhu and Singh, 2022; Raghavan and Khan, 1997).

Magnetorheological (MR) fluids are suspensions of magnetic particles which behave as a solid in the presence of a magnetic field due to the attractive dipolar interactions between the particles induced by the application of a magnetic field. Their rheological behavior is characterized by the presence of a yield stress whose value increases like $\sigma_y \sim H^{\alpha}$ with $\alpha$ close to 2 for low fields (typically H<50kA/m) when there is no saturation of the magnetisation. At intermediate fields the saturation of magnetisation of the particles in a small domain around the contact point gives $\alpha \sim 3/2$ (Ginder et al., 1996). At the highest fields (H~800kA/m) the magnetisation of the particles saturates , therefore the magnetic force and the yield stress saturate too and the maximum yield stress induced by the magnetic forces alone should remain below 100kPa. These predicted behaviors are quite well verified experimentally (Bossis et al., 2019; Fang et al., 2010).

In order to go beyond the limitation imposed by the saturation magnetisation, it is worth considering using a synergetic effect between the increase of viscosity induced by the application of a magnetic field and the one coming from the shear thickening behavior. One approach was to use a shear thickening fluid as the suspending medium of magnetic particles. Several authors have developed this approach based on a so-called MRSTF fluid. Most of them used silica nanoparticles - fumed silica - in ethylene or polyethylene glycol as STF together with CIP particles (Guo et al., 2023; Ortigosa Moya, 2021; Sokolovski et al., 2020; Zhang et al., 2008; Zhao et al., 2023). H.Wang et al (Wang et al., 2024), used a mixture of silica and sepiolite particles in polyethylene glycol together with CI particles; Qian et al (Qian et al., 2024) added boron nitride particles to fumed silica and CI particles for magnetic field-assisted polishing. In all cases it appears that adding CI particles decreases the shear thickening effect and there is no indication that it increases the MR effect compared to a suspension containing the same amount of CI particles, nevertheless the presence of silica nanoparticles has the noticeable effect to reduce the sedimentation of CI particles. Actually, there is no need to use a STF to get a shear thickening and even a DST in a MR suspension. With a suspension of CI particles in silicone oil (Jiang et al., 2015; Tian et al., 2010), a DST transition was observed with a large increase of the shear stress at a low shear rate (below $1s^{-1}$) by applying a large magnetic field (B>0.4 Tesla) on the suspension. The DST transition at high field was also obtained with a suspension of CI particles in sunflower oil (Chen et al., 2014) but the transition disappeared when the particles were coated with graphite oxide, demonstrating the importance of frictional contact to get the DST transition. We also have observed a DST transition in suspensions of CI particles in ethylene glycol, but only in the presence of a superplasticiser molecule which allows to reach a high enough volume fraction of CI particles, typically above 60%. In this case a small magnetic field (10-50kA/m) allows to trigger the DST transition and to obtain jumps of stresses above 100kPa (Bossis et al., 2022). It is also possible to use an STF made from a suspension of micronic particles like calcium carbonate (CC) or silica particles presenting a DST at shear stresses around 100Pa. Then adding a small fraction of CIP (5-10% in volume) is enough to trigger the DST transition with a small magnetic field and at the same time to decrease the sedimentation of CIP particles which are embedded inside the network of less dense CC particles (Bossis et al., 2023).

In our preceding work we have explored the effect of the magnetic field on the DST transition in a steady flow in relation with the associated jump of viscosity. Here we study the effect of the magnetic field on the DST transition in the oscillatory regime. We shall present results on the storage modulus G' and loss modulus G" and compare the change of modulus induced by the magnetic field to the one induced by the DST transition. We shall also show that the critical stress can depend on the frequency of oscillations and on the way that the magnetic field is applied to the suspension.

## II. MATERIALS AND METHODS

We have used two kinds of particles: the carbonyl iron (CI) particles (grade HQ from BASF) have a density of $\rho$=7.7g/cm$^3$ measured by gas pycnometer. The calcium carbonate (CC) particles (grade BL200 from Omya) have a density of $\rho$=2.72 g/cm3. Their average radii and standard deviation determined from image analysis of electronic microscopy pictures were respectively $R_{CI}$=0.297µm; $\sigma_{stdCI}$= 0.15µm and $R_{CC}$=1.75µm. $\sigma_{stdCC}$= 0.809µm. The suspending fluid is a mixture of polyethylene glycol (85%) and water (15%), this ratio minimizing the evaporation. The superplasticizer molecule is a diphosphonate-polyethylene oxide (commercial name Optima 100) used in cement industry and made of 44 O-$CH_2CH_2$ groups. The phosphonate groups at the end of the molecule have a negative charge compensated by a sodium counter ion. The sodium ions are replaced either by iron atoms from CI particles or calcium atoms from CC particles when the phosphonate head adsorbs onto the particles. The repulsive force between the particles is generated by the Brownian motion of the PEO chains which extend in the suspending fluid: when two particles are pushed against each other by the applied shear stress, the interpenetration of the PEO chains decreases their entropy resulting in a repulsive force. The critical stress, needed to locally expel the polymer layer from the surface of the particles and producing the DST transition, is very sensitive to the concentration of the polymer. The experimental results presented here were obtained with a concentration of wt=0.4% (weight of the dry polymer relative to the weight of the particles). The rheological measurements were realized on a MCR502 rheometer in plate-plate geometry with titanium disk of radius 2cm and a gap around 1mm. A drop of suspension was placed in the center of the tray and the upper disc was slowly lowered in rotation until the suspension reaches the entire edge of the disc To prevent wall slip, a sandpaper disk, of 60 μm average roughness, cut out with a punch, was stuck on both plates thanks to its adhesive face in order to prevent slipping at the interface. The absence of slip was verified using different gaps in both steady-state (Fig.1) and oscillatory flow (Fig.4). As can be seen in these figures, no significant variation or trend is observed as a function of the gap on the rheological curves which is evidence of absence of slippage both at 5Hz and in steady or large deformation.

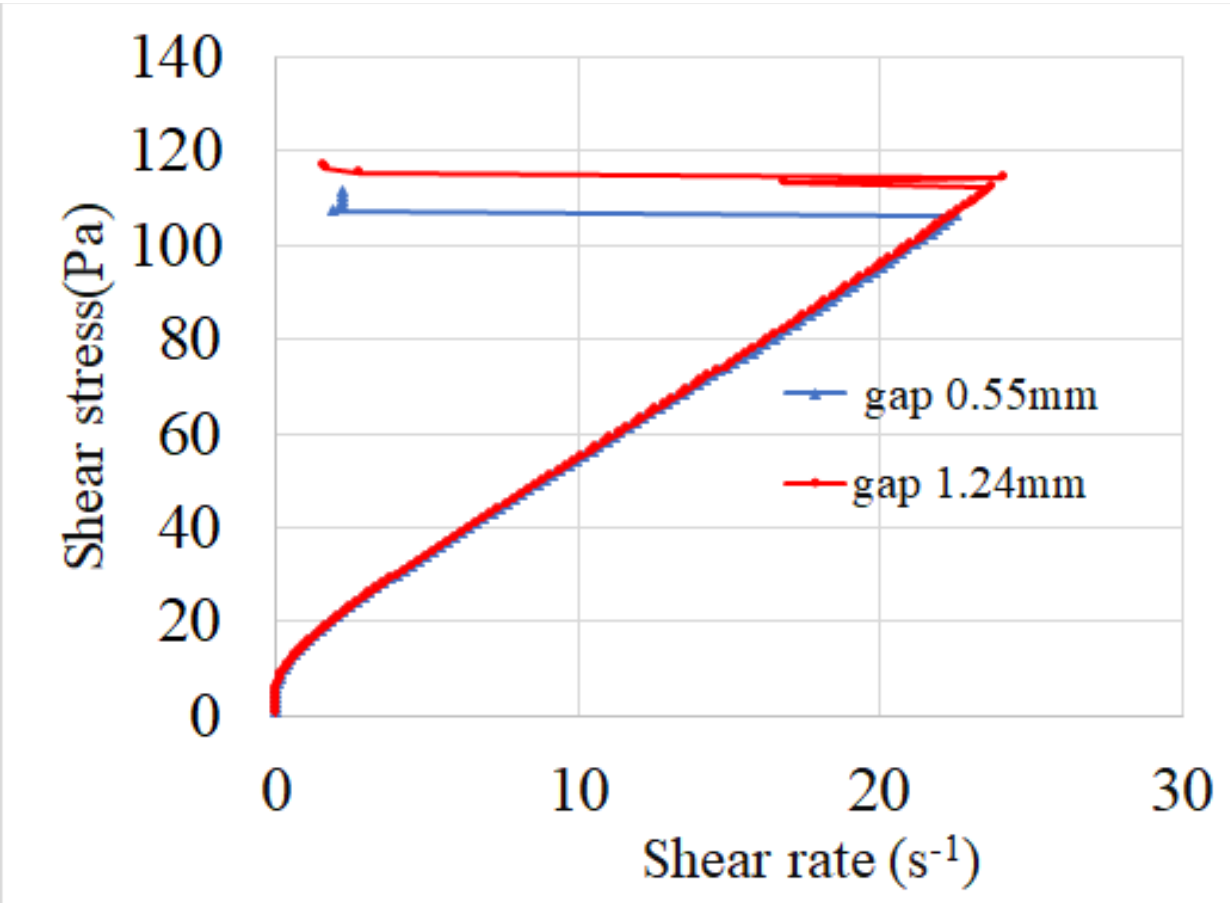

Fig.1 Stress versus shear rate at zero field for two different gaps of 0.55mm and 1.24mm. The volume fraction of CIP is Φ= 64%

The sample was positioned in the middle of a homemade coil of internal diameter 7.5cm which was cooled with water circulation around. The field H= 7.2kA/m at the middle of the coil corresponds to a current of 1A in the coil. The increase in axial field between the edge of the upper disc of radius 2cm and its center represents 4.5% of the field value at the center. This field gradient, as well as the gravity can induce a non-homogeneous density profile and it is worth giving an estimate of this effect. The magnetophoretic force on a particle is given by $\vec{F}_m = \mu_0 v_p \chi_p H \vec{\nabla} H$ and the gravity force by $\vec{F}_g = (\rho_p - \rho_f) v_p \vec{g}$ where $\chi_p$ (~30-50 for iron) is the initial susceptibility , $v_p$ the volume of the particle and $\mu_0=4\pi 10^{-7}$ the vacuum permeability. The velocity of an isolated particle is given by the Stokes law: $\vec{V_0} = \vec{F_p}$ /(6π $\eta_f$ R). With $\eta_f$~0.01Pa.s the viscosity of the suspending fluid and R~0.5μm the average radius of the particles. For a field H=30kA/m and a field gradient $\vec{\nabla}$H =0.045H/0.02 we obtain $V_{m0}$=1.57μm/s for the radial velocity due to the field gradient. The velocity induced by gravity being $V_{g0}$= 0.9μm/s. These are velocities of particles alone in the fluid; the velocity inside a suspension of volume fraction Φis quite well represented by the Richardson-Zaki (Richardson and Zaki, 1954) expression V= $V_0(1-\Phi)^n$ with n≈5. For Φ=0.6 we have V= 0.01 $V_0$, therefore, in the characteristic time of 3 minutes for the measurement with an applied field, the displacement of the particles remains negligible on the order of a few microns. Furthermore, after each experiment, a 3-minute pre-shear at $10 s^{-1}$ was performed to homogenize the suspension, and no drift in the results was observed after a series of experiments. The temperature of the sample was maintained at 20°C thanks to a Peltier thermostat below the bottom plate and the measured viscosity of the suspending fluid at this temperature was $\eta_f$=0.012 Pa.s.

## III. DST IN STEADY SHEAR FLOW

Before dealing with the viscoelasticity behavior induced by the application of a magnetic field in the presence of DST we will present some characteristic behaviour of these fluids in steady flow conditions. In Fig.2 we have the stress versus

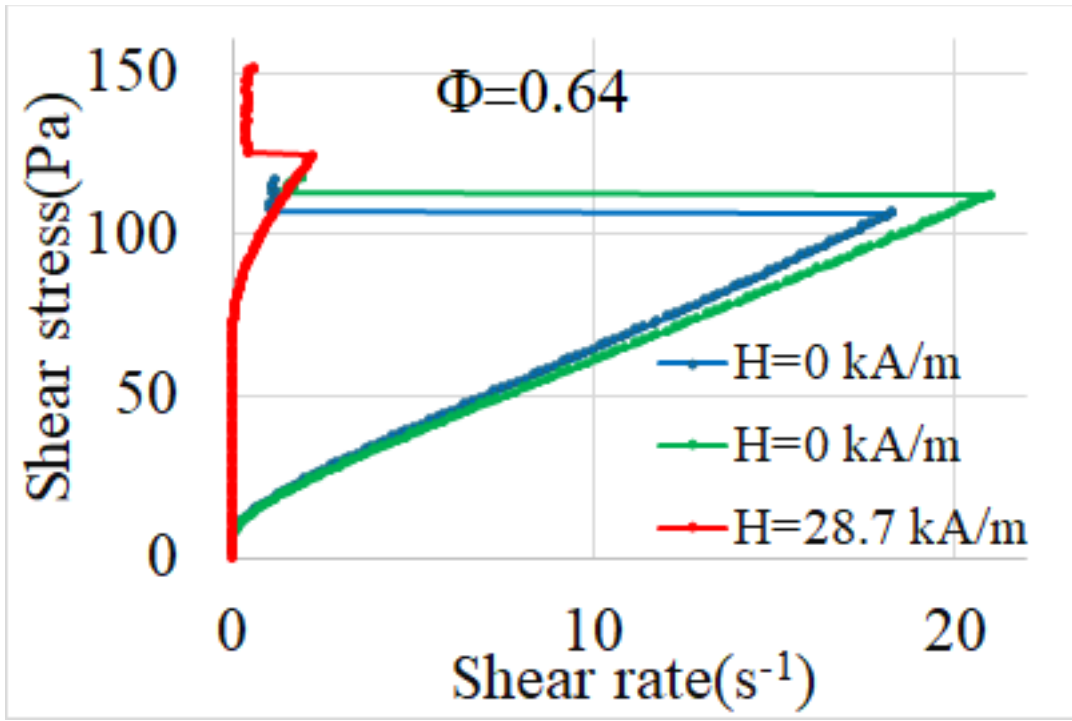

Fig.2 Stress versus shear rate in stress-imposed mode without magnetic field: green and blue curves and with a magnetic field H=28.7kA/m: red curve. The volume fraction of CIP is Φ= 64%

shear rate for a suspension of carbonyl iron particles at a volume fraction of 64% in ethylene glycol/water suspension with 0.4% wt of superplasticiser relatively to the weight of CIP particles. A ramp of stress was applied with a rising time of 1Pa/s. In the absence of field, we have a stress that we can perfectly fit ($R^2$=0.9998 for both curves) with a Bingham law: $\sigma = \sigma_y + \eta_B \dot{\gamma}$ between $\dot{\gamma} = 2\ s^{-1}$ and the critical shear rate $\dot{\gamma}_c = 18.3\ s^{-1}$ (the corresponding critical stress is $\sigma_c$=106.3Pa). The parameters of the fit are the dynamic yield stress $\sigma_y$=14.6 Pa and the Bingham viscosity: $\eta_B$=4.97 Pa.s. The second curve at zero field was made on a new suspension and gives an idea of the reproducibility of the experiments; here the fit gives $\sigma_y$=14.8 Pa, $\eta_B$=4.6 Pa.s and at the transition we have $\dot{\gamma}_c = 21\ s^{-1}$ and $\sigma_c$=112Pa.

It is important to note that the DST transition happens abruptly without being preceded by a shear thickening. The theory proposed by Wyart and Cates (Wyart and Cates, 2014), introduces a function 0 < f(σ/$\sigma_c$) <1 describing the increase of the proportion of frictional contacts between particles. The general shape of this function is a sigmoid, but here we have a Heaviside function meaning that at the critical stress ,$\sigma_c$, the proportion of frictional contacts passes abruptly from zero to one. We have interpreted this behavior based on the fact that, in a very concentrated suspension, once the stress is high enough to break the coating layer of polymer between two particles they form a quasi-solid aggregate which will increase the shearing stress between this aggregate and a neighbouring particle and so on. If this process is fast enough compared to the time needed to separate the particles (typically $1/\dot{\gamma}$) then we have this "instantaneous" jamming associated with a sharp decrease of the shear rate as shown in Figs.1,2. The growth in the size of frictional aggregates, that leads to percolation capable of blocking the flow, also depends on the volume fraction and requires increasingly higher critical stresses as the concentration decreases. In practice, for these carbonyl iron suspensions, the blocking transition disappears completely for $\Phi < 0.54$ and at $\Phi$=0.54 the critical stress is larger than 1000 Pa (Bossis et al., 2022). We should expect that this jamming completely blocks the flow but it is not the case, and the suspension keeps flowing with a small shear rate. We have proposed a model which reproduces this behaviour (Bossis et al., 2022) based on the relaxation of the frictional contacts once the shear rate decreases since in the absence of flow the polymer can adsorb again in the contact region and separate the particles. If we apply a magnetic field (red curve in Fig. 2), first, as expected, we have an increase of the yield stress and secondly a strong decrease of the critical shear rate from 21 $s^{-1}$ to 2.3 $s^{-1}$. This decrease can be qualitatively explained by the fact that the magnetic field produces a strong increase of the viscosity and the critical stress needed to overcome the repulsive force should remain approximately the same as without field. Here there is a small difference of critical stress between H=0 and H=28.7 kA/m (~20 Pa) but the hydrodynamic force is maximum on the compression axis at 45° of the rotational axis whereas the magnetic force is maximum at 0°, therefore, we cannot simply add these two forces together. We shall come back to this point in the next section devoted to the oscillatory regime, but before it is useful to present characteristic results in steady shear obtained with a suspension made of admixture of magnetic (CIP) and non-magnetic ($CaCO_3$) particles. It is represented in Fig.3 for the mixture 10% CIP and 90% $CaCO_3$ in volume and $\Phi$=0.69. The weight of polymers relative to the total weight of particles is still 0.4%. The first thing to note is that, if we impose a ramp of shear rate instead of a ramp of stress, then when the critical shear rate is exceeded, we have a jump of stress instead of a decrease of the shear rate at the DST transition (cf. Fig.2). This difference in behavior arises from the fact that the rheological curve presents a reversal point in the shear rate when the stress is increased; therefore, if instead of imposing the stress, we impose the shear rate to go beyond this point, we have a stress jump whose amplitude will only be limited by fractures in the percolated aggregate or slips at the walls if these are smooth at the particle scale.

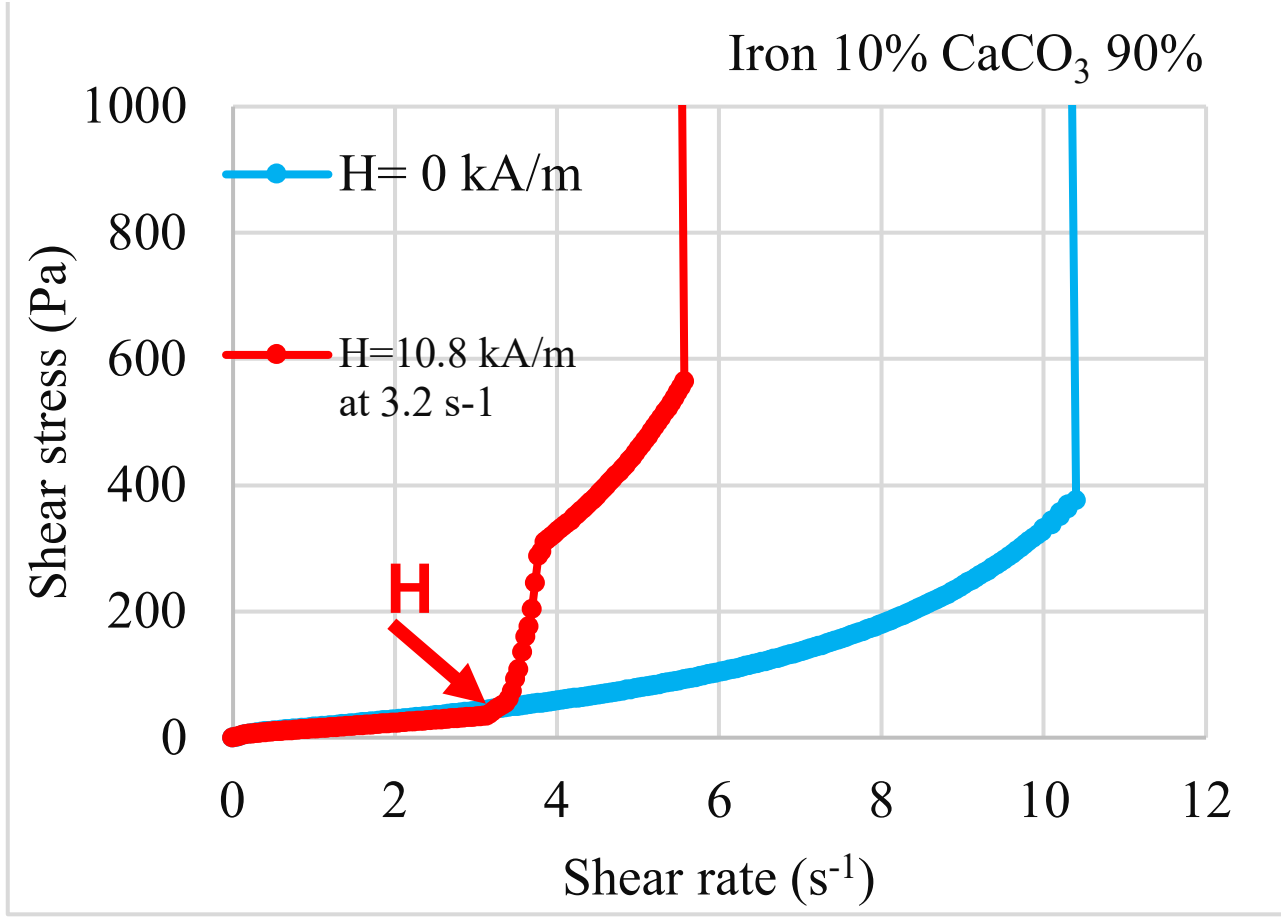

Fig.3 Stress versus shear rate in shear -imposed mode without magnetic field: blue curve, and with a magnetic field H=28.7kA/m raised progressively from 3.2 $s^{-1}$: red curve. The volume fraction of CIP is 10% of the total volume of particles and the total volume fraction of particles is $\Phi$=69%

Since this jump is larger than the one corresponding to the maximum torque of the rheometer: 0.23N.m, we stop the record after the jump. We see that the mixture $CaCO_3$/CIP presents a DST transition at a critical stress ($\sigma_c$~380Pa) much larger than the pure CIP system at $\Phi$=0.64 ($\sigma_c$~110Pa); this is because the effective volume fraction of $CaCO_3$ is lower: $\Phi_{CC}= V_{CC}/V_{tot}$ =0.60 and the critical stress increases when the volume fraction decreases. For the red curve we have begun to progressively increase the field at a shear rate of 3.2 $s^{-1}$. The increase of stress is the one corresponding to the dynamic yield stress, here 270Pa for a field of 10.8 kA/m. This field is kept constant and the shear rate continues to increase; now the jump of stress takes place at 5.6 $s^{-1}$ for a stress of 564Pa instead of respectively 10.4 $s^{-1}$ and 376 Pa at zero field. As for the case of pure CIP system the critical stress increases in the presence of the magnetic field and the difference-here 188Pa- due to the attractive interaction induced by the magnetic field is again smaller than the dynamic yield stress. Finally, we see that 10% of magnetic particles is enough to trigger the DST transition of 90% of non-magnetic particles. This synergetic effect, where small aggregates of CIP provokes under shear the percolation of frictional contacts between magnetic and non-magnetic particles, was predicted by numerical simulations (Wilson and Klingenberg, 2017) and was described in details taking into account the difference in size between the two kinds of particles (Bossis et al., 2023). Whereas the effect of transition for magnetorheological fluids in steady shear flow is rather well documented, this is not the case in the oscillating regime which is the subject of the next section.

## IV. DST TRANSITIONS IN OSCILLATING SHEAR FLOW AT ZERO FIELD

Before presenting the viscoelastic moduli, we first focus on comparing the critical stresses associated with the blocking transition in steady state and in oscillatory regime. Simultaneously, we ensured that there was no slippage at the walls by using different gaps. The results presented in Fig. 4. Show that they are well reproducible and do not depend on the gap. From these results, the standard deviation of the critical stress is $\Delta\sigma_c/\sigma_c=\pm7.5\%$ and that of the critical strain is: $\Delta\gamma_c/\gamma_c= \pm5.5\%$. The average value is $\sigma_c=86.6\pm6.5$Pa, significantly lower than in steady motion where $\sigma_c=110\pm8$Pa

As briefly recalled in the introduction most of the studies on shear thickening magnetorheological fluid (MRSTF) are made by adding carbonyl iron particles (CIP)

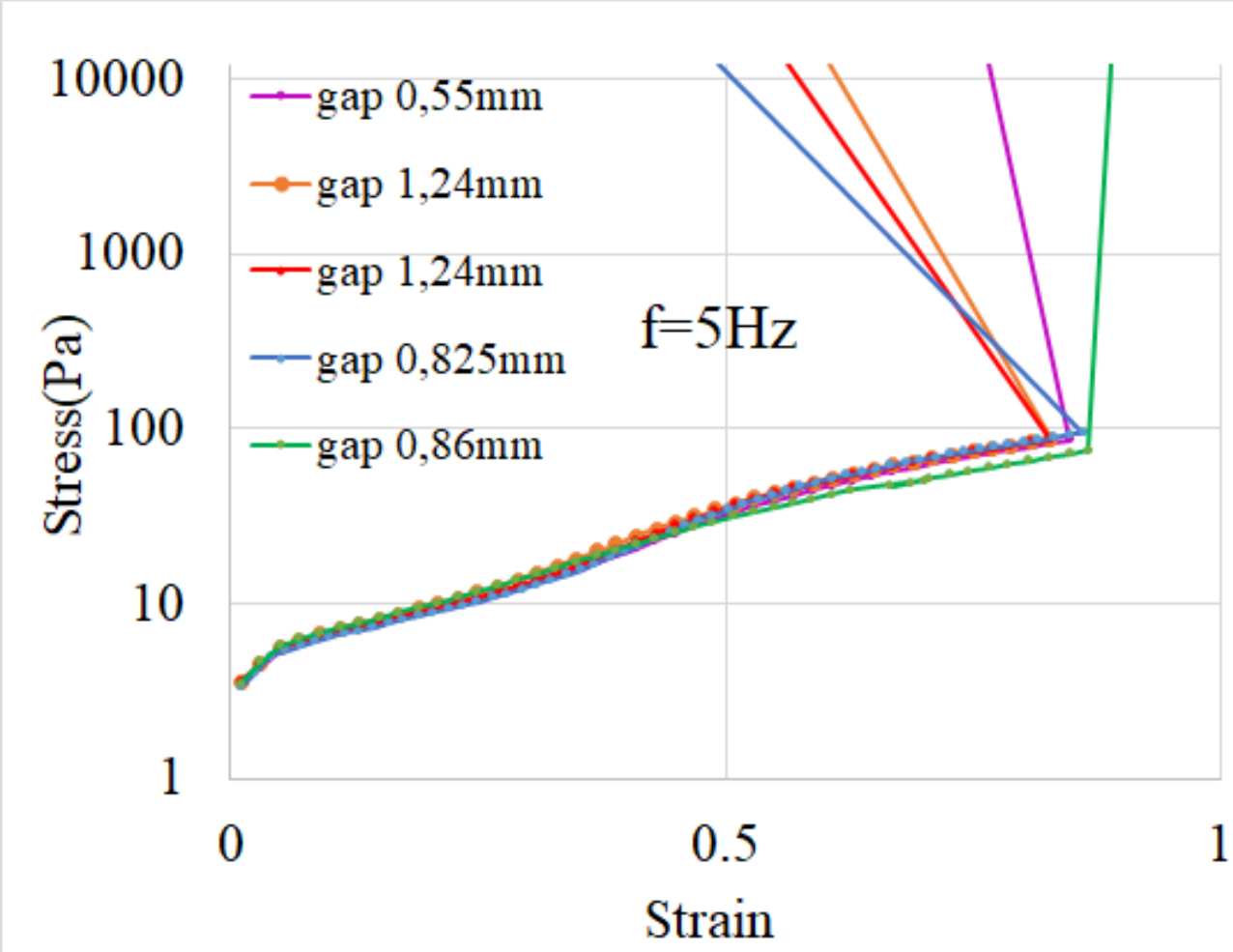

Fig.4 Stress versus strain at f=5Hz for different gaps during a ramp of strain

to a shear thickening suspension made of silica nanoparticles (fumed silica) dispersed in ethylene glycol or polyethylene glycol. The silica nanoparticles have a size between 10 and 20nm but they are present as fractal aggregates with a typical size of a few hundred nanometers (Raghavan and Khan, 1997). If these suspensions with a volume fraction of silica between 10 and 40% present a shear thickening effect in the absence of magnetic field, this effect disappears progressively when an increasing magnetic field is applied which is disappointing for the applications where a synergetic effect between the shear thickening and the magnetic field was expected. In our suspensions we will see that we keep this synergetic effect, but before we will present a few results on the viscoelastic behavior in the absence of the field. Typical curves displaying the storage, G', and loss, G'', moduli versus strain for different frequencies are shown respectively in Fig.5 and Fig.6. The first important thing to note is that we have a strong discontinuity of the two moduli above a critical strain $\gamma_c$. In particular, at the frequency of 5 Hz, the jump of stress corresponds to a torque larger than the maximum torque of the rheometer (0. 23N.m) and the strain is no longer regulated, therefore we did not use these values above the transition; the values of G', G'' reported in table I are those below the transition. For the lower frequency,0.1Hz, the ramp of strain was stopped at $\gamma=6$ but we still did not observe the DST transition. A second experiment allowed to reach the transition at a strain $\gamma_c=27.4$

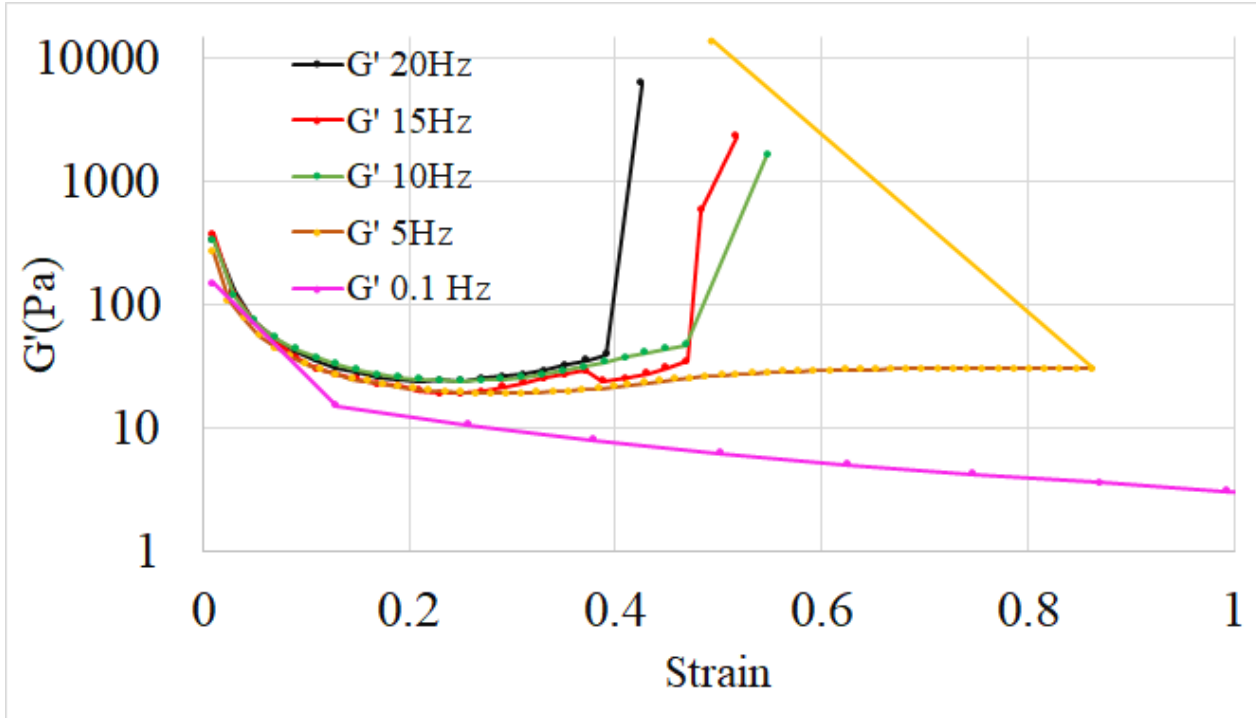

Fig.5. Storage modulus, G', versus strain for different frequencies. CIP particles $\Phi=0.64$

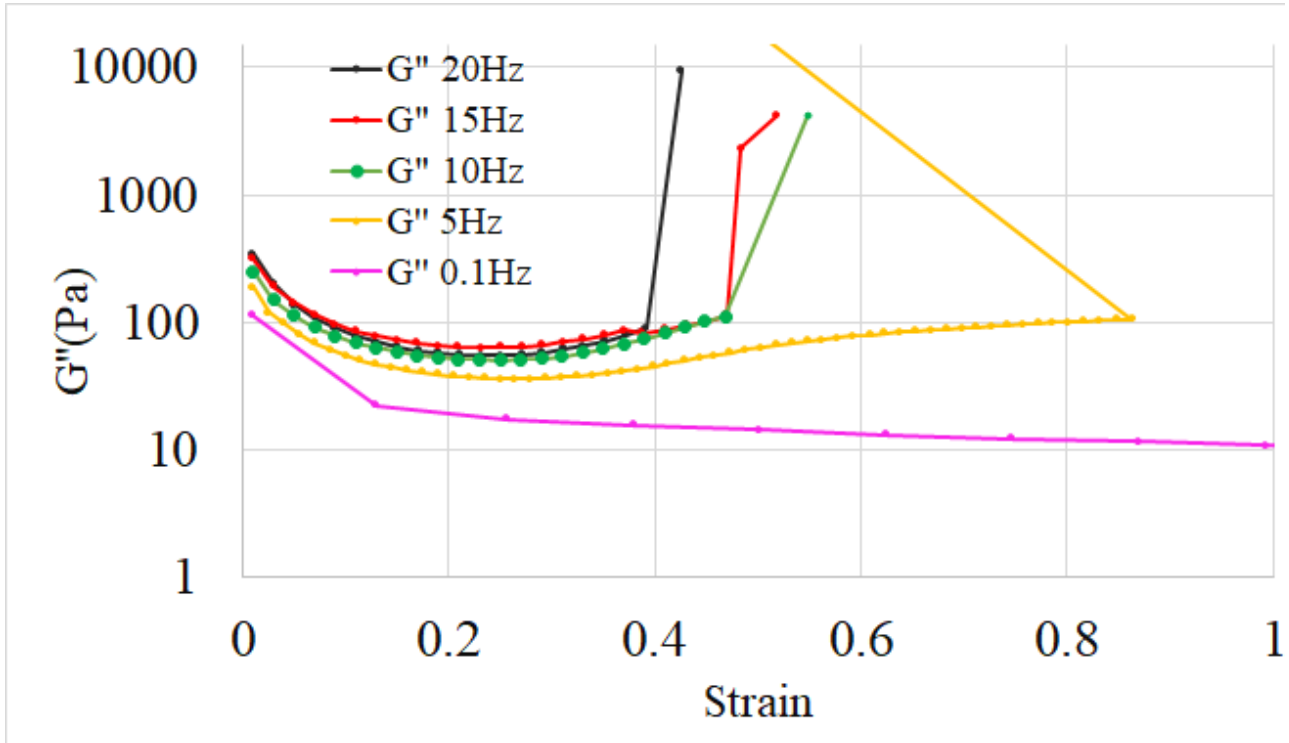

Fig.6 Loss modulus, G'', versus strain for different frequencies. CIP particles $\Phi=0.64$.

The loss modulus is always larger than the storage modulus, except at very low strain (not visible on these figures); the crossover being at a strain between 0.01 and 0.02 for all the frequencies. At strains smaller than 0.01 the behavior is the one of a gel with a yield stress, which can be seen in Fig.1, and then we have a shear thinning associated with the progressive rupture of this gel until the third region of shear thickening begins for $\gamma>0.3$. except for 0.1Hz where G' is shear thinning up to the transition. The sudden blockage appears at a critical strain $\gamma_c=0.86$ at 5Hz; it is largely reduced for f=10Hz with $\gamma_c= 0.47$ and asymptotes at higher frequency ($\gamma_c=0.41$ at 20Hz). On the other hand, the critical stress continuously decreases with the frequency from 96Pa at 0.1 Hz to 38Pa at 20Hz. We know that the DST happens for a critical stress which generates a compressive force between particles larger than the repulsive one, here due to the adsorbed polymer. Assuming that this critical stress is constant whatever the frequency and taking the dynamic shear rate $\dot{\gamma} = \gamma.\omega$ we have $\gamma_c=\sigma_c/(\eta.\omega)$.Then if $\eta$ does not depend on the frequency we expect that $\gamma_c \propto 1/\omega$. To take into account the plateau at high frequency it was proposed to use $\gamma_c=\gamma_\infty + \sigma_c/(\eta.\omega)$ (Raghavan and Khan, 1997). The presence of this plateau was explained by a sticking probability between aggregates which should decrease as $1/\omega$ (Laun et al., 1991), or by the presence of a wall slip with $\gamma_\infty=2\ D_{slip}/h$ (Lee and Wagner, 2003) but in our case the two disks are covered with sand paper to prevent slippage of the suspension and we have seen, using different gaps, that there was no slip. To help the understanding of the evolution of the moduli with the strain and the frequency it is usual to compare the viscosity obtained in the steady shear flow to the complex viscosity: $\eta^* = (\sqrt{G'^2 + G''^2}\,)/\omega \sim \sigma_{dyn}/(\gamma\omega)$ where $\sigma_{dyn}$ is the amplitude of the oscillating stress. This complex viscosity for f=0.1Hz, f=5 Hz, f=20Hz is compared to the steady one in Fig.7

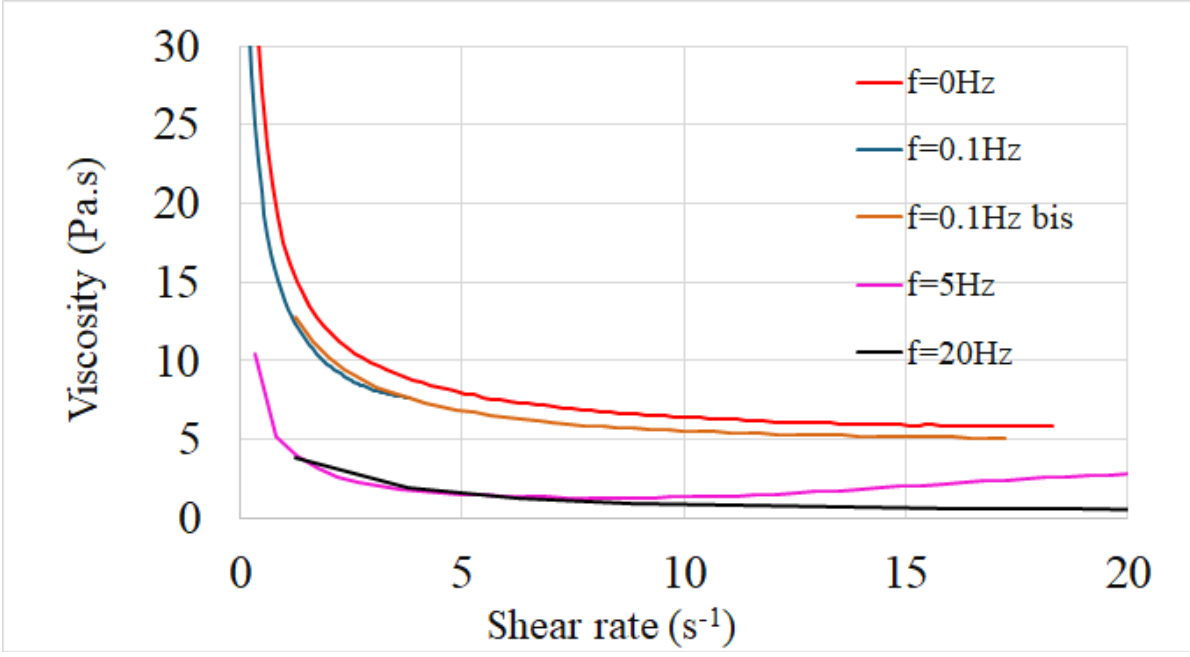

Fig.7 Steady viscosity (red curve f=0Hz) versus $\dot{\gamma}$ and dynamic viscosity at 0.1Hz,5Hz and 20 Hz versus γω. The curve at 0.1Hz was recorded in two different experiments.

We see that the steady viscosity $\sigma = \eta/\dot{\gamma}$ (red curve) is the highest one and that the dynamic viscosity obtained at 0.1 Hz is close whereas the others at 5 and 20Hz have a much lower viscosity. Actually, at f=0.1 Hz the maximum strain of the curve corresponds to $\gamma$=27.4, and all the points corresponding to viscosities smaller than 18 Pa.s correspond to strains larger than unity. On the contrary for the curves at f=5Hz or f=20Hz all the points correspond to strain smaller than unity. If the strain is smaller than unity we have a "cage effect" that is to say each particle can move in the cage formed by its neighbours without being obliged to escape. In this situation the particles are not pushed in close contact and that minimizes the viscous dissipation. The decrease of the viscosity with the frequency does not come from the frequency itself but from the fact that the attainable strain before the jamming is lower than unity. It also explains the difference of viscosity between the frequencies 5Hz and 20Hz above 10 $s^{-1}$ because at f=5Hz the maximum strain before jamming is $\gamma_c$=0.86 and we have already $\gamma$=0.64 for $\gamma\omega$=20$s^{-1}$ so we are approaching $\gamma$=1 faster than at f=20 Hz where $\gamma$=0.16 still far from $\gamma_c$=0.41. In conclusion we can say that at low frequency, so at high strain, the viscosity does not depend on frequency, and we recover the 1/ω behavior of the critical strain; then at higher frequency the strain becomes smaller than unity and the cage effect decreases progressively the viscosity which can explain the plateau for $\gamma_c$ if this decrease is proportional to 1/ω. Nevertheless if we continue to increase the frequency, the viscosity will likely stop to decrease at strains smaller than about $\gamma$=0.3 and we can expect to find another law for $\gamma_c$ versus the frequency. Another observation is that the critical stress, $\sigma_c$, just before the transition decreases strongly with the frequency since we have $\sigma_c$ =110 Pa in steady flow, $\sigma_c$= 86.6 Pa at 5 Hz and $\sigma_c$=38Pa at 20Hz. A possible explanation could be related to the dynamic of adsorption of the polymer on the surface of the particles: if the period of oscillation becomes shorter than the characteristic time of desorption of the polymer it can prevent the formation of the coating layer and so decrease the stress needed to bring the particles in contact.

## V. DST TRANSITIONS IN OSCILLATING SHEAR FLOW TRIGGERED BY A MAGNETIC FIELD

We are mainly interested to see how the application of a magnetic field modifies the DST transition in the oscillatory regime. We also want to compare the change of modulus induced by the magnetic field alone to the change of modulus induced by the field in the presence of the DST transition.

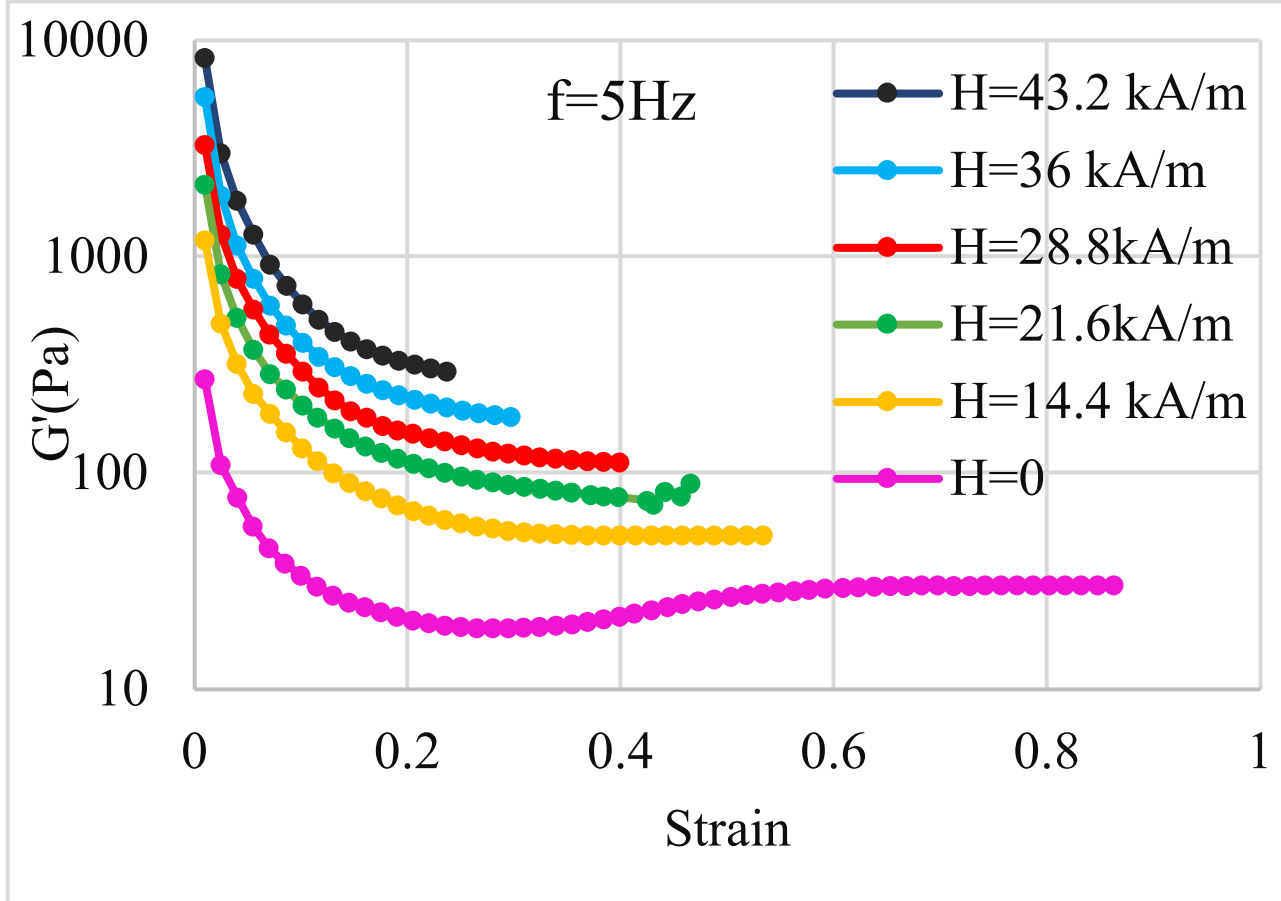

Fig.8 Storage modulus versus imposed strain at a frequency f=5Hz in presence of different magnetic fields; the last point of each curve corresponds to the DST transition

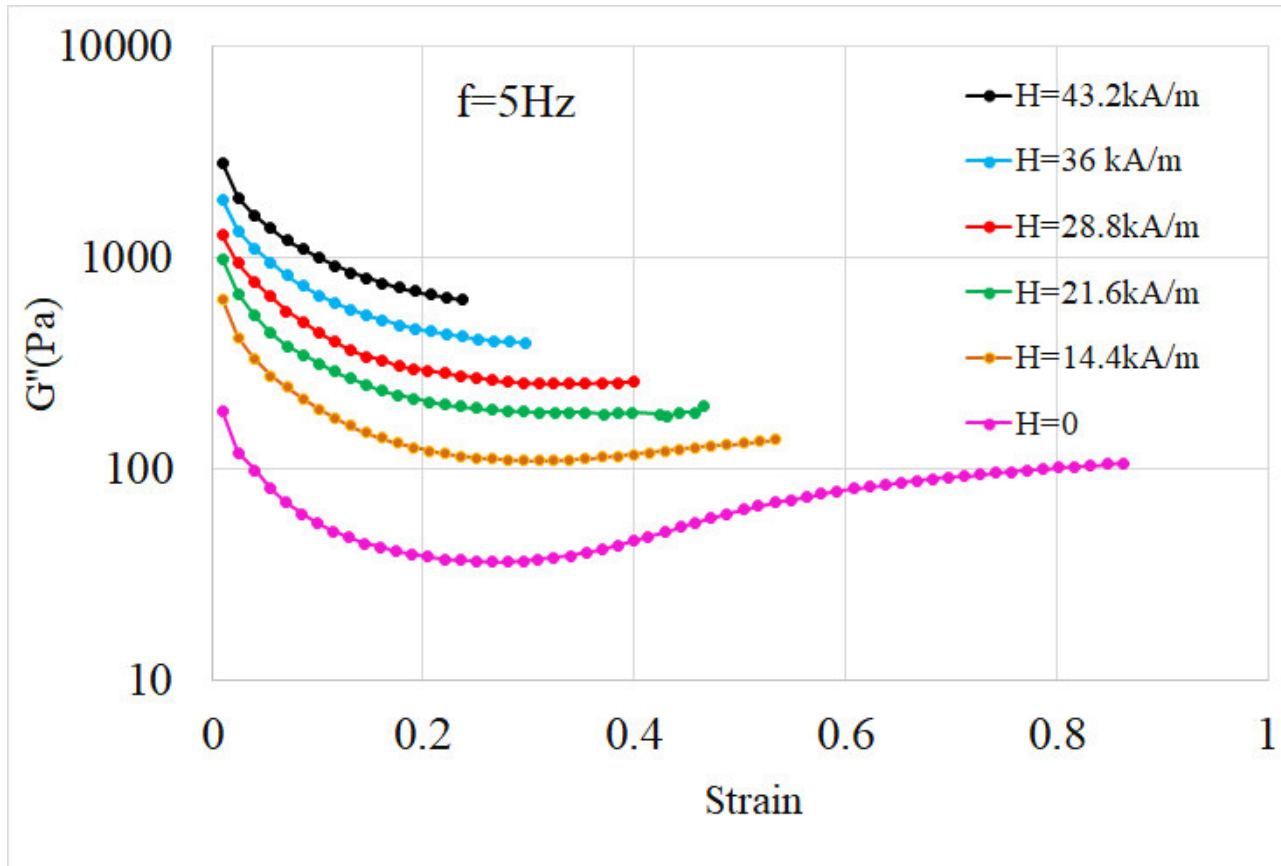

Fig.9 Loss modulus versus imposed strain at a frequency f=5Hz in presence of different magnetic fields; the last point of each curve corresponds to the DST transition

One way to do that is to compare the situation where the field is turned on before beginning a ramp of strain to the situation where the field is turned on during a ramp of strain. The first case is presented in Figs. 8 and 9 for different fields. For the clarity of the figures we did not represent the points above the transition, so the last point is the one just before the jamming transition. We see the magnetorheological effect with an important increase of the storage and loss moduli in the presence of the field. For instance, at a strain $\gamma$=0.2 we pass from G'=20Pa to G'=304 Pa for H= 43.2kA/m, but for a strain of 0.01 we pass from G'=270 to G'=8300Pa, that is to say a ΔG'~8000Pa compared to 280Pa at $\gamma$=0.2. This difference comes from the increase in the separation of particles with the strain and the fact that the local magnetic field decreases strongly with the separation. On the other hand, except at the lower strain, $\gamma$=0.01, the loss modulus is always larger than the storage modulus, meaning that for larger strain we have a viscoelastic liquid. In the following table we have given the moduli for three values of the strain: 0.01,0.1,0.2 and the critical values $\gamma_c$, $\sigma_c$ of the jamming transition for the five values of the field that we have used. We note again that the critical stress at zero field, 86.6Pa is significantly lower than the one in steady flow $\sigma_c$= 110 Pa

| H(kA/m) | 0 | 14.4 | 21.6 | 28.8 | 36 | 43.2 |
|---|---|---|---|---|---|---|
| $\gamma_c$ | 0.86 | 0.53 | 0.47 | 0.4 | 0.3 | 0.24 |
| $\sigma_c$ | 86.6 | 83 | 107 | 112 | 128 | 165 |
| $G'_{0.01}$ | 270 | 1189 | 2157 | 3287 | 5464 | 8293 |
| $G''_{0.01}$ | 188.5 | 638.6 | 983 | 1279 | 1881 | 2817 |
| $G'_{0.1}$ | 33 | 130 | 205 | 293 | 399 | 602 |
| $G''_{0.1}$ | 55.1 | 191.6 | 314.2 | 440.8 | 663.5 | 998 |
| $G'_{0.2}$ | 20.7 | 66.6 | 110 | 151 | 217 | 316 |
| $G''_{0.2}$ | 38.1 | 121.2 | 207.6 | 291.5 | 445.9 | 670 |

Table 1 Frequency f=5Hz. Values of the moduli G' and G'' for three values of strain: $\gamma$=0.01,$\gamma$=0.1,$\gamma$=0.2. The two first lines correspond to the critical strain and critical stress at the jamming transition.

Several models have been proposed to predict the increase of shear or loss modulus due to the application of a magnetic field. Among the models based on the evolution of the chain structure with the strain we can distinguish between the model of Jolly et al (Jolly et al., 1996) and Ginder et al (Ginder et al., 1996), that take into account the saturation magnetisation of the regions of particles which are in contact; the last model predicting for low field a storage modulus proportional to the field $G'=3\mu_0\Phi M_s H$, where $\mu_0$ is the magnetic permeability of vacuum and $M_s$ the saturation magnetisation of the magnetic material. The other models, either using a multipolar development to calculate the magnetostatic force between two particles (Klingenberg and Zukoski, 1990; López-López et al., 2005) or simply the dipolar approximation, predict a storage modulus growing as the square of the magnetic field. A fit of the increase of shear modulus $\Delta G(H)= G'(H)-G'(H=0)$ versus H with a power law:$\Delta G(H) \propto H^n$ gives n=1.94 for a strain $\gamma$=0.01, n=1.56 for $\gamma$=0.1 and n=1.65 for $\gamma$=0.2. These theories for the storage modulus only apply at the lowest strain $\gamma$=0.01 where G'>G''. With an exponent of 1.94 for $\gamma$=0.01, it suggests that, at low field, the saturation magnetisation does not play a significant role in this very dense suspension. The reason can be that, due to the high volume fraction $\Phi$=0.64 the field inside the suspension $H = H_0/\mu_r$ is very low since $\mu_r$, the relative permeability of the suspension, is around 6 at low field; cf. Table 1 in (Bossis et al., 2019). Another, more original observation that can be drawn from Figs 8 and 9 is that the critical deformation associated with jamming can be greatly reduced by increasing the field (cf. Table 1). It means that we can easily produce a large jump of storage and loss moduli thanks to the triggering of friction by a magnetic field. Actually, here the transition is not triggered since the field is present from the start, we simply observe its shift as a function of the field's amplitude. In the following figures we present the jump of the moduli when the field is applied during a ramp of strain which is the situation of interest for the damping applications.

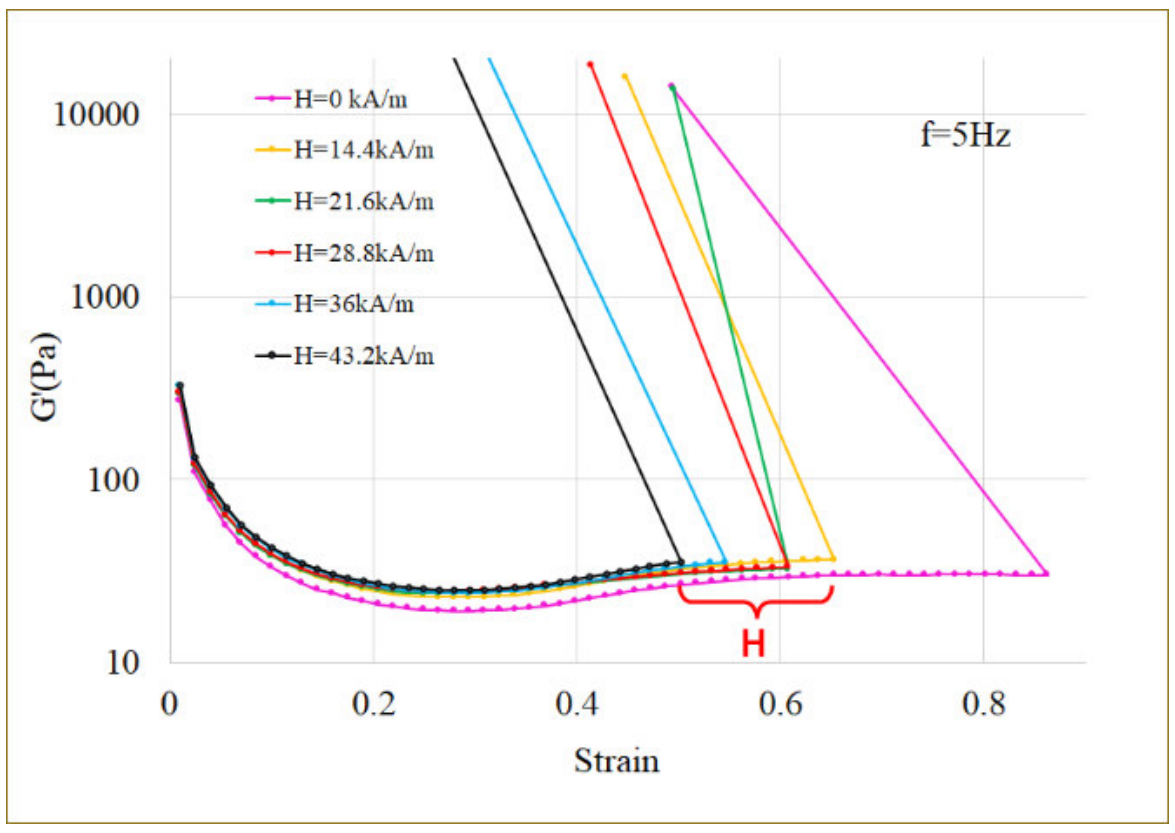

Fig.10 Frequency f=5Hz. Application of the magnetic field at different strains between $\gamma$=0.65(yellow) to $\gamma$=0.5 (black). The curve at H=0 (purple) is the reference one at $\gamma_c$=0.86

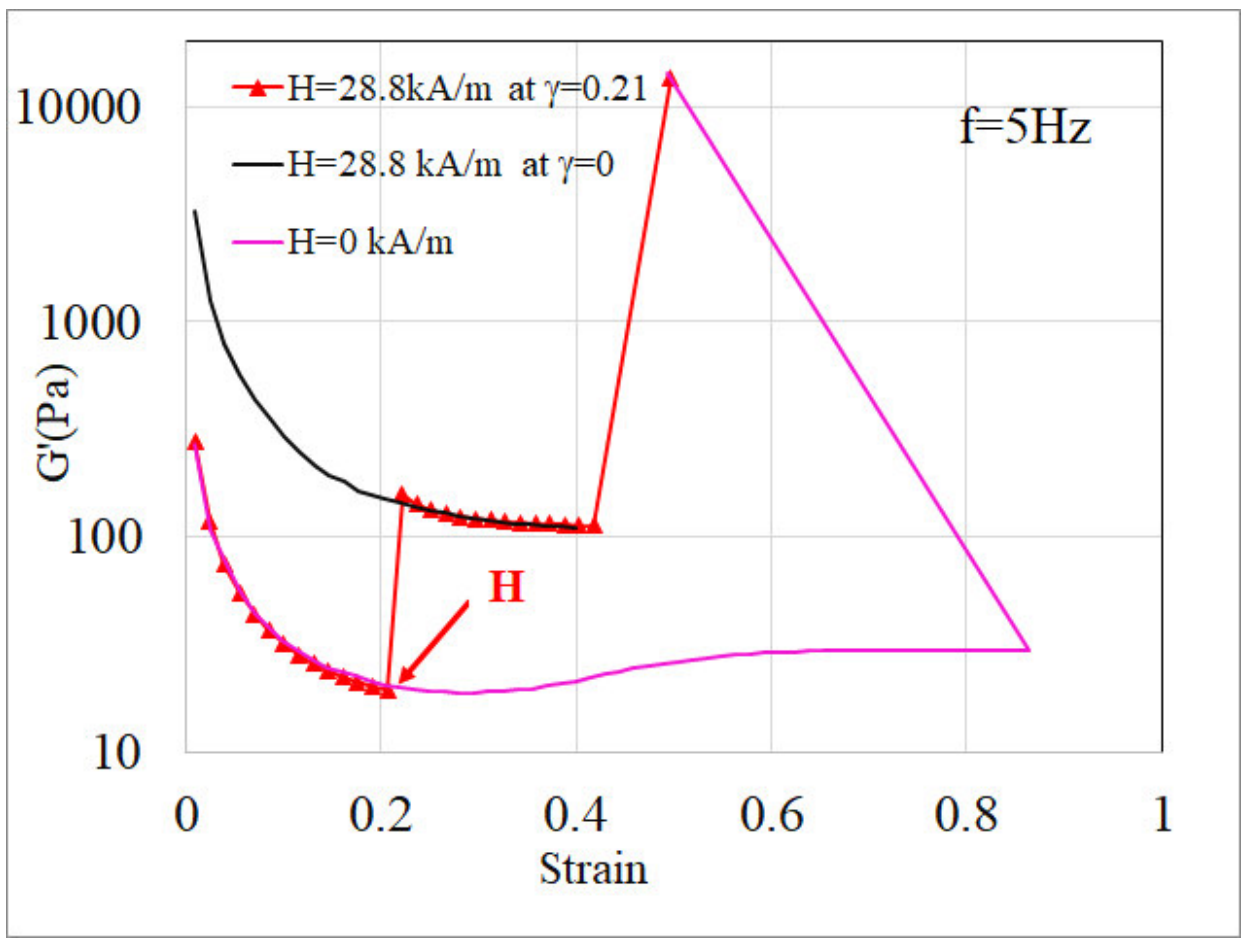

Fig.11 Frequency f=5Hz. Purple curve H=0. Red curve the field H=28.8 kA/m is applied at $\gamma$=0.21. Black curve: the same field H=28.8 kA/m is applied from the beginning ($\gamma$=0)

All the curves are made one after the other; the reference one is without field (H=0). For the following ones the field is applied at increasing strain except for H=21.6kA/m and H=28.8 kA/m where the field is turned on at the same strain: $\gamma$= 0.6. For each field, the strain when we applied the field was always larger than the critical strain obtained when the field was applied from the beginning (cf. $\gamma_c$ table 1) so it immediately provokes the jamming transition. On the contrary if we apply the field at a strain smaller than the critical one corresponding to this particular field, then we only have the jump of modulus bringing to the curve measured with the field imposed from the beginning. This is exemplified in Fig.11 where we applied a field H=28.8 kA/m at a strain of 21% well below the critical strain $\gamma_c$=0.4 for this field (red curve); we only have a jump $\Delta G'$ =124 Pa joining the black curve which is the one corresponding to the same field but imposed from $\gamma$=0. So, on the one hand we have a modest increase of modulus ΔG' ~100Pa if we apply the field at a strain below the critical one but an increase ΔG'> 10000Pa if we apply it above $\gamma_c$ (cf. Fig. 10). Once again we recall that the values of G' above the transition should not be considered as the true ones first because of the limiting torque of 0.23N.m (corresponding to a stress $\sigma_{max}$=12250Pa with our geometry) and secondly because the displacement is no longer regulated so the determination of the phase shift,δ, cannot be trusted. In any event the order of magnitude of the jump of moduli for H=28.8 kA/m is just obtained by $\sigma_{max}/\gamma$ = 20080Pa where γ=0.61 is the strain when we have applied the field. In practice we see that triggering the friction with the magnetic field represents a gain of more than two orders of magnitude compared to what we can obtain solely with the magnetic field. The last point that needs to be considered is the dependence of the critical stress, $\sigma_c$, on the field for the different situations presented in the last figures. In Fig. 12 we have a plot of the critical stress versus the strain for the situation where the field is applied from the beginning (upper curve) like in Figs 8 and 9, or during the ramp of strain for the values of strain corresponding to those represented in Fig.10

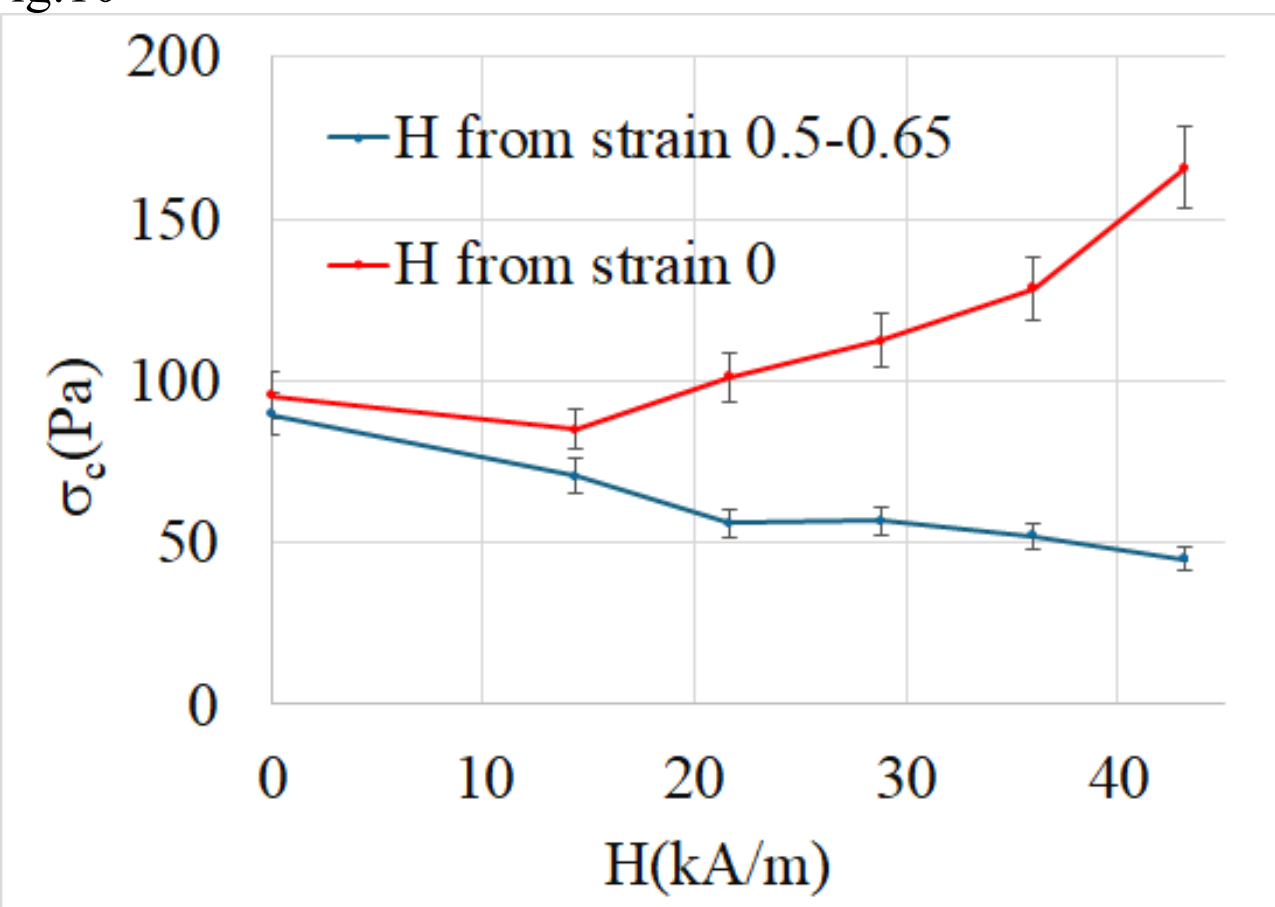

Fig.12 Critical stress versus the magnetic field. Either the field is applied from the beginning of the ramp of strain (red curve) or during the ramp (blue curve) at strains 0.5<γ<0.65. Oscillation frequency: f=5Hz. The error bars correspond to the uncertainty $\Delta\sigma_c$=±7.5% $\sigma_c$

The striking observation is that the two situations are fundamentally different since, when the field is applied from the beginning, the critical stress increases with the field whereas if it is applied at intermediate strain it decreases with the field. As discussed in the section on steady flow, the critical stress should be considered as the one which is able to overcome the repulsive force coming from the coating layer of the particles and should include both the hydrodynamic stress and the magnetic stress. Both the magnetic force and the hydrodynamic force depend on the structuration of the particles under the shear flow and the magnetic field. Under the magnetic dipolar force, $F_m$, the particles are structured in linear chains parallel to the direction of the field. The stability of these chains under an oscillating shear flow depend on the Mason number which expresses the ratio of shear stress to magnetic stress between two particles: $M_n = \eta_f \gamma\omega/(\mu_0\beta^2H^2)$ where $\beta=(\mu_p-\mu_f)/(\mu_p+2\mu_f)$. The magnetic permeability, $\mu_p$, of the iron particle is much larger than the one ,$\mu_f$ , of the suspending fluid so we can take β=1. For a field H=10kA/m, a frequency f=5Hz , a strain γ=1 and a viscosity $\eta_f$= 0.012Pa.s we have Mn~0.001. This low value, even for the lowest field, means that the chain structure is conserved during the oscillating shear flow as shown by numerical simulations ((Kang et al., 2012), Fig.13). If the magnetic field is applied from the beginning, this aligned structure is conserved during the oscillating flow. For particles aligned with the velocity gradient there is no axial hydrodynamic force between the particles, so the total applied stress needs to be larger in order to overcome the repulsive force that prevents the particles to enter in frictional contact. On the contrary in the absence of the field at already high shear rate ($\dot{\gamma} = \gamma.\omega > 15s^{-1}$ with γ=0.5 and f=5Hz), the structure is disordered with particles separated by the presence of the oscillating flow. When the field is applied in the presence of this already disordered suspension, the particles will not form compact aligned aggregates but rather isotropic aggregates giving rise to an increase of the hydrodynamic stress which, added to the magnetic stress can overcome the repulsive force at a smaller total applied stress than for compact aligned aggregates. This difference of structuration of the particles between the two cases is obviously the reason of this divergence between the two curves observed in Fig.12, but only numerical simulations like those described in (Seto et al., 2013) could help to validate our interpretation.

## VI. CONCLUSION

In this work we have presented experimental results concerning the viscoelastic moduli of magnetorheological suspensions made of carbonyl iron particles dispersed in ethylene glycol and coated by a superplasticiser molecule. Looking at the complex viscosity it appears that at f=0.1Hz it is close to the viscosity of the steady flow but much lower for frequencies above 5Hz. This evolution can be explained by the cage effect since from 5 Hz the strain is lower than unity, meaning that the deformation of the cage formed by the neighbours of a given particle is small, hence a lower viscous dissipation. In the absence of magnetic field, the moduli show a jump at a critical stress which is slightly lower than the one observed in steady flow. On the other hand, the critical stress decreases when the frequency increases. An explanation could be that the period of the oscillation becomes close to the residence time of a polymer on the surface of the particles. In the presence of a magnetic field the critical strain decreases when the field increases because the magnetic force contributes to the total compressive force between particles. The application of the magnetic field at a strain large enough to trigger the DST transition produces a jump of moduli whose amplitude is two orders of magnitude larger than the one produced by the magnetic field alone in absence of the DST transition (cf Fig.11). The major interest is that these large jumps are produced with low values of the magnetic field which is interesting for damping applications. At last, for a given frequency, we observe an increase of the critical stress when the field is applied from the beginning at zero strain whereas it is the contrary if the field is applied directly at strains larger than 0.5 (cf Fig.12). A tentative explanation is given based on the difference of structure induced in each case.

These results can be summarized through the dependence of the critical strain on the magnetic field and frequency:$\gamma_c=\sigma_c/(\omega\eta_{dyn})$ with $\eta_{dyn}$ shown in Fig.7 for H=0. For a practical application, we need to have an ensemble of experimental curves giving $\gamma_c$=f(H,ω) in order to be able to trigger the jump of the viscoelastic moduli when the strain overcomes a chosen value. Note also that the range of operability at low viscoelastic moduli with $\gamma<\gamma_c$ (H=0,ω) will be larger if we lower the volume fraction of the suspension but at the detriment of the amplitude of the jump of viscoelastic moduli for $\gamma>\gamma_c$. Numerous studies demonstrate the benefits of using shear-thickening fluids for vibration damping (Liu et al., 2023; Wei et al., 2019; Zhou et al., 2016) but not with shear-thickening magnetorheological fluids (MRSTF) because, as mentioned in the introduction, these fluids have a shear thickening that decreases with the applied field. This is not the case with our fluid but a too abrupt shear-thickening transition, such as that observed at a volume fraction of 64%, could be detrimental to the mechanical device. However, the advantage of MR fluids exhibiting a blocking transition lies in the ability to regulate the intensity of this transition by adjusting both the particle concentration and the intensity of the applied magnetic field. As discussed at the end of section II the sedimentation velocity is low at these high concentrations (of order 0.01μ/s) but over a long period the formation of a dense cake of height h at the bottom of a device produces a force~$\rho ghR^2$ whose value can exceed the repulsive force from the coating polymer and lead to irreversible aggregation Another challenge is designing a hydraulic circuit that minimizes variations in shear gradients. These gradients are responsible for the migration of particles into areas of low shear rate, and this increased concentration can also lead to irreversible aggregation. These problems of irreversible aggregation are nevertheless common to any kind of MR suspension and it is the amplitude of the repulsive barrier formed by the coating layer that can prevent them.

In this work, we were unable to properly quantify the jump in shear and loss moduli because it exceeds the capabilities of our rheometer. A study using a dynamic mechanical analysis (DMA) device would allow reliable measurements of viscoelastic moduli above the transition in a configuration close to that of a vibration damper.

## ACKNOWLEDGMENT

The authors want to thank the CENTRE NATIONAL D'ETUDES SPATIALES (CNES, the French Space Agency) for having supported this research.

## REFERENCES

Barnes, H.A., 1989. Shear-Thickening (“Dilatancy”) in Suspensions of Nonaggregating Solid Particles Dispersed in Newtonian Liquids. Journal of Rheology 33, 329–366. https://doi.org/10.1122/1.550017

Bossis, G., Ciffreo, A., Grasselli, Y., Volkova, O., 2023. Analysis of the rheology of magnetic bidisperse suspensions in the regime of discontinuous shear thickening. Rheol Acta 62, 205–223. https://doi.org/10.1007/s00397-023-01388-x

Bossis, G., Grasselli, Y., Volkova, O., 2022. Discontinuous Shear Thickening (DST) transition with spherical iron particles coated by adsorbed brush polymer. Physics of Fluids 34, 113317.

Bossis, G., Volkova, O., Grasselli, Y., Ciffreo, A., 2019. The Role of Volume Fraction and Additives on the Rheology of Suspensions of Micron Sized Iron Particles. Front. Mater. https://doi.org/10.3389/fmats.2019.00004

Chen, K., Zhang, W.L., Shan, L., Zhang, X., Meng, Y., Choi, H.J., Tian, Y., 2014. Magnetorheology of suspensions based on graphene oxide coated or added carbonyl iron microspheres and sunflower oil. Journal of Applied Physics 116, 153508.

Comtet, J., Chatté, G., Niguès, A., Bocquet, L., Siria, A., Colin, A., 2017. Pairwise frictional profile between particles determines discontinuous shear thickening transition in non-colloidal suspensions. Nature communications 8, 1–7.

Egres, R.G., Wagner, N.J., 2005. The rheology and microstructure of acicular precipitated calcium carbonate colloidal suspensions through the shear thickening transition. Journal of rheology 49, 719–746.

Fall, A., Bertrand, F., Hautemayou, D., Méziere, C., Moucheront, P., Lemaitre, A., Ovarlez, G., 2015. Macroscopic discontinuous shear thickening versus local shear jamming in cornstarch. Physical review letters 114, 098301.

Fang, F.F., Choi, H.J., Choi, W.S., 2010. Two-layer coating with polymer and carbon nanotube on magnetic carbonyl iron particle and its magnetorheology. Colloid Polym Sci 288, 359–363. https://doi.org/10.1007/s00396-009-2164-3

Franks, G.V., Zhou, Z., Duin, N.J., Boger, D.V., 2000. Effect of interparticle forces on shear thickening of oxide suspensions. Journal of Rheology 44, 759–779.

Frith, W.J., d’Haene, P., Buscall, R., Mewis, J., 1996. Shear thickening in model suspensions of sterically stabilized particles. Journal of Rheology 40, 531–548. https://doi.org/10.1122/1.550791

Ginder, J.M., Davis, L.C., Elie, L.D., 1996. Rheology of magnetorheological fluids: models and measurements. International journal of modern physics b 10, 3293–3303.

Guo, F., Xu, Z., Gu, J., 2023. Effects of nano-fumed silica and carbonyl iron powder of different particle sizes on the rheological properties of shear thickening fluids. Colloid Polym Sci 301, 539–555. https://doi.org/10.1007/s00396-023-05087-0

Jiang, J., Hu, G., Zhang, Z., Meng, Y., Tian, Y., 2015. Stick-slip behavior of magnetorheological fluids in simple linear shearing mode. Rheologica Acta 9–10, 859–867. https://doi.org/10.1007/s00397-015-0877-4

Jolly, M.R., Carlson, J.D., Munoz, B.C., 1996. A model of the behaviour of magnetorheological materials. Smart materials and structures 5, 607.

Kang, T.G., Hulsen, M.A., den Toonder, J.M., 2012. Dynamics of magnetic chains in a shear flow under the influence of a uniform magnetic field. Physics of Fluids 24.

Klingenberg, D.J., Zukoski, C.F., 1990. Jan. 1990. Studies on the steady-shear behavior of electrorheological suspensions. Langmuir 6, 15.

Larsen, R.J., Kim, J.-W., Zukoski, C.F., Weitz, D.A., 2014. Fluctuations in flow produced by competition between apparent wall slip and dilatancy. Rheologica Acta 53, 333–347.

Laun, H.M., 1994. Normal stresses in extremely shear thickening polymer dispersions. Journal of non-newtonian fluid mechanics 54, 87–108.

Laun, H.M., Bung, R., Schmidt, F., 1991. Rheology of extremely shear thickening polymer dispersionsa)(passively viscosity switching fluids). Journal of Rheology 35, 999–1034.

Lee, Y.S., Wagner, N.J., 2003. Dynamic properties of shear thickening colloidal suspensions. Rheol Acta 42, 199–208. https://doi.org/10.1007/s00397-002-0290-7

Liu, H., Fu, K., Cui, X., Zhu, H., Yang, B., 2023. Shear thickening fluid and its application in impact protection: a review. Polymers 15, 2238.

Lootens, D., Hébraud, P., Lécolier, E., Van Damme, H., 2004. Gelation, shear-thinning and shear-thickening in cement slurries. Oil & gas science and technology 59, 31–40.

López-López, M.T., De Vicente, J., Bossis, G., González-Caballero, F., Durán, J.D.G., 2005. Preparation of stable magnetorheological fluids based on extremely bimodal iron–magnetite suspensions. J. Mater. Res. 20, 874–881. https://doi.org/10.1557/JMR.2005.0108

Mari, R., Seto, R., Morris, J.F., Denn, M.M., 2014. Shear thickening, frictionless and frictional rheologies in non-Brownian suspensions. Journal of Rheology 58, 1693–1724. https://doi.org/10.1122/1.4890747

Morris, J.F., 2025. Dilatancy and pressures in suspensions. Journal of Rheology 69, 223–234.

Neuville, M., Bossis, G., Persello, J., Volkova, O., Boustingorry, P., Mosquet, M., 2012. Rheology of a gypsum suspension in the presence of different superplasticizers. Journal of rheology 56, 435–451.

Ortigosa Moya, E.M., 2021. On the rheology of shear-thickening and magnetorheological fluids under strong confinement (PhD Thesis). Universidad de Granada.

Prabhu, T.A., Singh, A., 2022. Rheology and microstructure of discontinuous shear thickening suspensions. Journal of Rheology 66, 731–747.

Qian, C., Tian, Y., Ahmad, S., Ma, Z., Li, L., Fan, Z., 2024. Theoretical and experimental investigation on magnetorheological shear thickening polishing force using multi-pole coupling magnetic field. Journal of Materials Processing Technology 328, 118414.

Raghavan, S.R., Khan, S.A., 1997. Shear-thickening response of fumed silica suspensions under steady and oscillatory shear. Journal of colloid and interface science 185, 57–67.

Richardson, J.F., Zaki, W.N., 1954. The sedimentation of a suspension of uniform spheres under conditions of viscous flow. Chemical Engineering Science 3, 65–73.

Seto, R., Mari, R., Morris, J.F., Denn, M.M., 2013. Discontinuous shear thickening of frictional hard-sphere suspensions. Physical review letters 111, 218301.

Sokolovski, V., Tian, T., Ding, J., Li, W., 2020. Fabrication and characterisation of magnetorheological shear thickening fluids. Frontiers in Materials 7, 595100.

Tian, Y., Jiang, J., Meng, Y., Wen, S., 2010. A shear thickening phenomenon in magnetic field controlled-dipolar suspensions. Applied Physics Letters 97, 151904.

Wang, H., Du, C., Ge, M., Sun, L., Fu, Y., 2024. Dynamic excellence in magnetorheological shear thickening fluid: Fabrication and formulation of a constitutive model. Physics of Fluids 36.

Wei, M., Huang, Z., You, A., Sun, L., 2025. Application of shear-thickening fluid in impact resistance: a review. Bull Mater Sci 48, 102. https://doi.org/10.1007/s12034-025-03468-5

Wei, M., Lin, K., Guo, Q., Sun, H., 2019. Characterization and performance analysis of a shear thickening fluid damper. Measurement and Control 52, 72–80. https://doi.org/10.1177/0020294018819543

Wilson, B.T., Klingenberg, D.J., 2017. A jamming-like mechanism of yield-stress increase caused by addition of nonmagnetizable particles to magnetorheological suspensions. Journal of Rheology 61, 601–611.

Wyart, M., Cates, M.E., 2014. Discontinuous shear thickening without inertia in dense non-Brownian suspensions. Physical review letters 112, 098302.

Zhang, X., Li, W., Gong, X.L., 2008. Study on magnetorheological shear thickening fluid. Smart materials and structures 17, 015051.

Zhao, Q., Wang, K., Yuan, J., Jiang, H., 2023. Research on rheological properties and phenomenological theory-based constitutive model of magnetorheological shear thickening fluids. Smart Materials and Structures 32, 105037.

Zhou, H., Yan, L., Jiang, W., Xuan, S., Gong, X., 2016. Shear thickening fluid–based energy-free damper: Design and dynamic characteristics. Journal of Intelligent Material Systems and Structures 27, 208–220. https://doi.org/10.1177/1045389X14563869